%% file: main.tex
\documentclass[%
 reprint,
 amsmath,amssymb,
 aps,prl,
]{revtex4-2}
\usepackage{braket}
\usepackage{enumitem}
\usepackage{float}
\usepackage{graphicx}% Include figure files
\usepackage[english]{babel}
\usepackage{dcolumn}% Align table columns on decimal point
\usepackage{bm}% bold math
\usepackage{tikz}
\usetikzlibrary{calc, arrows.meta, fadings, shadings}
\usepackage{quantikz}

\usepackage{amsthm}
\usepackage{xcolor}
\usepackage{comment}

\newtheorem{theorem}{Theorem}
\newtheorem{definition}[theorem]{Definition}

\newtheorem{corollary}[theorem]{Corollary}

\DeclareMathOperator{\Tr}{Tr}

\begin{document}

\title{Demystifying Objectivity with Operator Algebra Quantum Error Correction}

% Force line breaks with \\

\author{Marin Girard}
\email{mcgirard@vt.edu}
% \affiliation{%
% Department of Physics, Virginia Tech, Blacksburg, VA, USA 24061\\
% Virginia Tech Center for Quantum Information Science and Engineering, Blacksburg, VA 24061, USA
\author{Gong Cheng}
\email{gongc@vt.edu}
% \affiliation{%
% Department of Physics, Virginia Tech, Blacksburg, VA, USA 24061\\
% Virginia Tech Center for Quantum Information Science and Engineering, Blacksburg, VA 24061, USA

%  \homepage{http://www.Second.institution.edu/~Charlie.Author}
% \affiliation{
%  Second institution and/or address\\
%  This line break forced% with \\
% }%
% \affiliation{
%  Third institution, the second for Charlie Author
% }%
\author{ChunJun Cao}
\email{cjcao@vt.edu}
\affiliation{%
Department of Physics, Virginia Tech, Blacksburg, VA, USA 24061\\
Virginia Tech Center for Quantum Information Science and Engineering, Blacksburg, VA 24061, USA
}%


\begin{abstract}
Quantum Darwinism extends the decoherence formalism to explain how objectivity emerges from quantum mechanics. However, existing approaches often capture only partial aspects of objectivity. By connecting quantum Darwinism to operator algebra quantum error correction, we show that the emergence of objectivity can be identified with the algebraic local recoverability of quantum codes. Applying this algebraic framework to stabilizer codes, we show that it yields a far more precise characterization of classicality and redundancy, unifies the traditional measures of objectivity, enables efficient classification via coding-theoretic tools, and supports large-scale Clifford simulations of decoherence dynamics.
\end{abstract}

\maketitle
\paragraph*{Introduction.}
An important open question in quantum physics is how classicality emerges from unitary quantum dynamics. The decoherence formalism provides part of the answer --- through its interaction with a vast environment, a coherent quantum system becomes entangled with environmental degrees of freedom that contain the measurement apparatus. The decohered quantum system is then approximately given by a classical mixture upon tracing out the environment. However, to recover a notion of objectivity where different observers can agree on the measurement outcome, decoherence alone is not enough.

\begin{figure}
    \centering
    \includegraphics[width=1\linewidth]{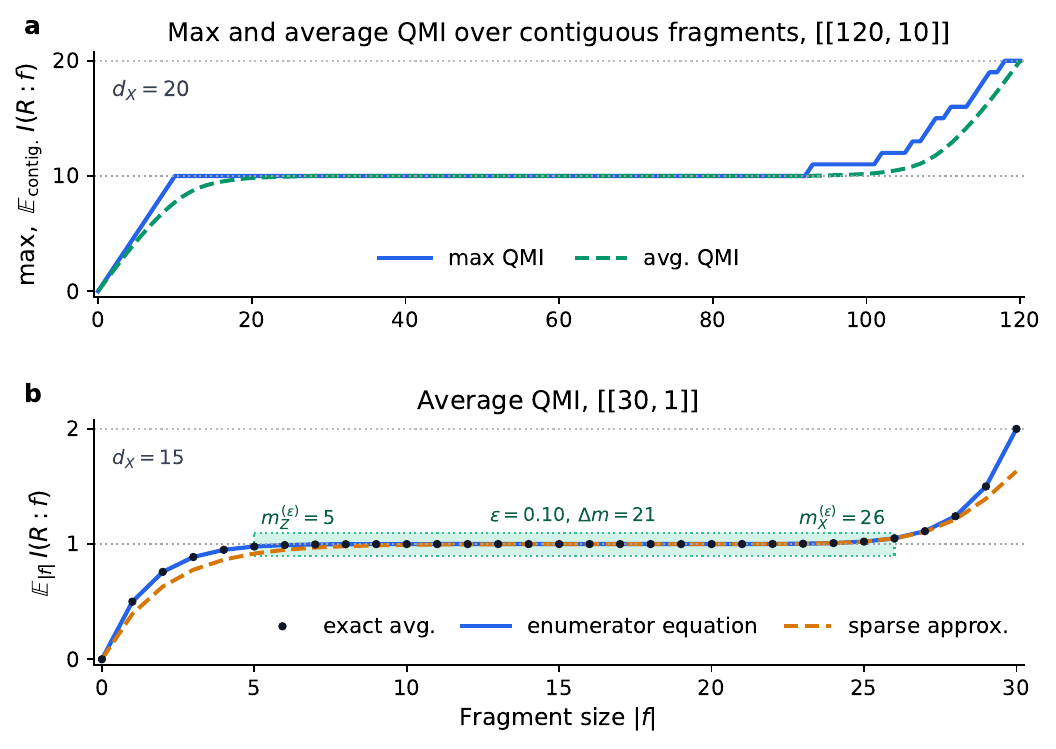}
    \caption{QMI plateaus in asymmetric CSS codes. a) $\max_{|f|}{I}(R:f)$ and $\mathbb{E}_{|f|}(R:f)$ for a $[[120,10]]$ code with $d_Z=1, d_X=20$. Maximization and averaging are done over contiguous fragments. b) $\mathbb{E}_{|f|}(R:f)$ for a $[[30,1]]$ code with $d_Z=1,d_X=15$, compared with (\ref{eqn:cosetavgQMI}) and the sparse approximation (\ref{eq:avg_QMI_exact}). Green lines mark the $\epsilon=0.1$ plateau bound of Corollary~\ref{cor:avg_plateau}.}
    \label{fig:QDQEC_plateaus}
\end{figure}

Quantum Darwinism (QD) addresses the emergence of objectivity by requiring information about a system  to be robustly and redundantly encoded in its environment, allowing independent observers to recover and compare the same record \cite{blume-kohout_simple_2005,blume-kohout_quantum_2006,zurek_quantum_2009}.
Existing approaches span a broad hierarchy: at one end, QD is often diagnosed by a ``classical plateau'' in the quantum mutual information (QMI) between the system and fragments of the environment, as illustrated in Fig.~\ref{fig:QDQEC_plateaus}; at the other end, spectrum broadcast structure (SBS) \cite{horodecki_quantum_2015} gives a much stronger condition on the joint system-environment state. Intermediate notions, such as strong quantum Darwinism, supplement these criteria with additional information-theoretic constraints \cite{le_strong_2019,korbicz_roads_2021}.

%What remains missing is a framework that identifies which information is objective: how much information a fragment contains, whether that information is classical or quantum, and how redundantly that same information is available. 
%\CC{the problem statement is still vague to me. Existing framework does determine how much info a fragment contains by computing QMI, and as you mentioned before, quantum discord was used to determine quantumness. It may not be refined, but it does do so. I will just propose something here and you can see if you agree. then we modify/refine? it sounds like a) the existing frameworks or approaches need to be more precise and b) computations are hard. For a) the information diagnostics are not super precise. This is especially apparent when dealing with multiple systems where not all of them are objective/classical.  For example, [subsequent text...] } \MG{My point was that no approach did all that together, but I wasnt clear. I agree with your suggestions}

These criteria capture important aspects of objectivity, but they leave two related gaps. First, they do not provide a unified way to compare how objective different observables are, nor do they specify how the redundant records are distributed across the environment. Second, current approaches are computationally expensive to evaluate in large models, but small-scale analysis alone may miss key features of objectivity.

The first limitation is apparent in systems with several relevant degrees of freedom, where different observables may become objective to different extents or in different fragments. In such cases, the averaged QMI can lack a plateau even as objective records exist. Moreover, because QMI includes both classical and quantum correlations, quantum discord may contribute to the apparent signal and spoof non-averaged QMI measures \cite{chisholm_importance_2024}. Conversely, SBS a la \cite{horodecki_quantum_2015} can exclude states that still contain operationally accessible redundant classical information. %\footnote{While this is true of SBS as described in the original paper \cite{horodecki_quantum_2015}, a more general form has been discussed in \cite{korbicz_roads_2021}, which does not suffer from this issue.}.

The second limitation is computational. Current models of QD and emergent objectivity, such as spin-chain and oscillator models of decoherence, can be technically challenging and numerically intensive \cite{ryan_quantum_2021,zurek_quantum_2022,balaneskovic_dissipation_2016}. Although recent constructions have made important analytic and numerical progress on QD--encoding transitions \cite{ferte_solvable_2024,ferte_decoherent_2025}, they are focused on tree-like architectures. Furthermore, computing von Neumann entropies and QMIs can grow exponentially with system size, ultimately limiting large-scale analysis. %their specificity and complexity can limit scalability and generality, making it difficult to isolate the essential ingredients for emergent objectivity.

In this letter, we formulate an algebraic approach to objectivity where classicality and redundancy of information, two key components of objectivity, are clearly defined and directly computable. 
We first establish a link between the encoding of quantum error correcting codes (QECCs) and decoherence --- the decohering quantum system corresponds to the logical data of a QECC while the system-environment interaction is represented by the encoding unitary of the code. Using known results of operator algebra QEC \cite{harlow_ryu-takayanagi_2017}, we show that the redundant classical records are precisely the redundantly represented commuting logical subalgebras that are supported on multiple disjoint fragments of the environment. Each fragment corresponds in the code picture to a subset of the physical qubits. This perspective leads us to a unified framework of algebraic objectivity, which incorporates many existing measures of QD and objectivity while providing a computable and quantitatively precise characterization of the proliferation and robustness of classical information.

This link also facilitates rigorous analysis of emergent objectivity by drawing upon many helpful structures of quantum coding theory. In particular, we show that asymmetry in code distance and the coset weight enumerator polynomials are connected to the existence of classical plateaus in the QMI diagnostics of QD. 

Focusing on stabilizer codes, which serve as minimal models for QD and decoherence, we introduce novel classes of QD models using classical linear codes and asymmetric quantum codes. By  composing these small components, we construct structurally rich but conceptually simple dynamical decoherence processes generated by local Clifford circuits where the emergent lightcones illustrate the proliferation of classical information and the loss of quantum coherence. These models are efficiently simulable thanks to the Gottesman-Knill theorem \cite{gottesman_stabilizer_1997} and can in principle be scaled to thousands of qubits.

\paragraph*{Decoherence, QECC, and operator algebra.}
A simple toy example of decoherence can be seen as follows. Initially, the system-environment state is $|\psi\rangle |e_0\rangle$ where a two-level system is in the state $|\psi\rangle=a|0\rangle+b|1\rangle$ and the environment is ancillary in the state $|e_0\rangle = |0\rangle^{\otimes n-1}$.

A quantum measurement takes place as a particular system-environment interaction $U_{SE}$ entangles the two systems, picking out $\{|0\rangle,|1\rangle\}$ as the pointer basis, such that the state approaches
\begin{equation}
    |\psi\rangle_S|e_0\rangle_E\xrightarrow{U_{SE}} |\tilde{\psi}\rangle=a|0\rangle_S|0\rangle^{\otimes n-1}_{E}+b|1\rangle_S|1\rangle^{\otimes n-1}_{E}
\end{equation}
in the long time limit. 
It is clear that $\rho_S =\Tr_E[|\tilde{\psi}\rangle\langle\tilde{\psi}|] = |a|^2|0\rangle\langle 0|+|b|^2|1\rangle\langle 1|$ becomes a classical mixture with outcomes consistent with a measurement in the $Z$ basis. 

Furthermore, the information about the system is also redundantly encoded in many fragments of the environment. Let $f$ be any subset of qubits in $E$ which is accessible to an observer, then
\begin{align}
    \rho_f &=|a|^2|00\dots\rangle\langle 00\dots|+|b|^2|11\dots\rangle\langle 11\dots|\\
    &= U_f \big[(|a|^2|0\rangle\langle 0|+|b|^2|1\rangle\langle 1|)\otimes \chi\big] U_f^\dagger,
    \label{eqn:ghzrecov}
\end{align} 
where the observer can recover a single classical bit of information about the system with $\chi$ being the tensor product of $|0\rangle\langle0|$ states. Therefore, distinct observers with access to different fragments all agree on the recovered classical information \cite{zwolak_redundancy_2017}.

If we compute the QMI $I(S:f)$ between the system $S$ and a fragment $f$, then as a function of the fragment size $|f|$, we see a classical plateau analogous to those shown in Fig.~\ref{fig:QDQEC_plateaus}, which is often treated as a key signature of emergent classicality and QD. 

A reader familiar with quantum error correction will quickly recognize that if we identify $S$ as the logical qubit and $E$ as the pre-encoding ancillary state, then $U_{SE}$ is nothing but the encoding unitary of an $[[n,1,1]]$ quantum repetition code with stabilizer group $\mathcal S = \langle Z_iZ_{i+1}; i=1,\dots, n-1\rangle$ and logical operators $\tilde{X}=X^{\otimes n}, \tilde{Z}=Z_1$. 

Since any logical operator multiplied by an element of the stabilizer group gives an equivalent physical representation, for any fragment with size $|f|<|SE|$, one can always find a representation of $\tilde{Z}$, but not $\tilde{X}$, supported in $f$. 

Under the unitary $U_{SE}$, the image of initial system Hilbert space $\mathcal{H}_S$  defines  the code subspace $\mathcal C$ while the joint system $\mathcal{H}_{SE}$ corresponds to the physical Hilbert space $\mathcal {H}_P$ of a QECC. In this model the original system $S$ is not a separate output subsystem, rather, its information is encoded across the joint degrees of freedom $SE$.

The fact that observers with qubits in $f$ can always implement a commuting logical subalgebra $M_f=\{\tilde{I},\tilde Z\}$ but not $\tilde X$ precisely reflects the fact that observers accessing various fragments can recover the same classical bit of the system but not the encoded quantum bit. 
Indeed, this observation generalizes to all  operator algebra quantum erasure correction codes, which include stabilizer codes. 

Let $\mathcal H_P=\mathcal{H}_f\otimes \mathcal{H}_{\bar f}$ and $M_f\subseteq A_{\rm logical}$ be the logical von Neumann subalgebra accessible on $f$ with $M_f'$ being its commutant. From the classification theorem, the algebra induces a (Wedderburn) decomposition of the Hilbert space $\mathcal C = \oplus_\alpha (\mathcal H_{f^\alpha}\otimes \mathcal H_{\bar f^\alpha})$ such that $M_f=\oplus_\alpha(L(\mathcal{H}_{f^\alpha})\otimes I_{\bar f^\alpha})$, $M_f'=\oplus_\alpha(I_{ f^\alpha}\otimes L(\mathcal{H}_{\bar f^\alpha}))$, and the center of the algebra is $Z(M_f)=M_f\cap M_f' = \oplus_\alpha c_\alpha I_{f^\alpha}\otimes I_{\bar f^\alpha}$. Here $L(\mathcal{H})$ denotes the set of linear operators over $\mathcal{H}$ and $c_\alpha$'s are constants.

\begin{theorem}[Harlow's erasure theorem \cite{harlow_ryu-takayanagi_2017}]
\label{thm:harlow}
     Let $\{|\widetilde{\alpha,ij}\rangle\}$ be an orthonormal basis for $\mathcal{C}$ that is compatible with the Wedderburn decomposition induced by $M_f$. Then the following are equivalent: 
\begin{enumerate}
    \item For any operator $\tilde{O}\in M_f$ and any state $|\tilde{\psi}\rangle\in \mathcal {C}$, there exists an operator $O_f$ on $\mathcal{H}_f$ such that $\tilde{O}|\tilde{\psi}\rangle = O_f|\tilde{\psi}\rangle$ and $\tilde{O}^\dagger|\tilde{\psi}\rangle = O_f^\dagger|\tilde{\psi}\rangle$.
    \item We can decompose $\mathcal H_f = \oplus_\alpha (\mathcal H_{f^\alpha_1}\otimes \mathcal H_{f_2^\alpha})$ such that for all basis states $|\widetilde{\alpha, ij}\rangle$ there exists unitary transformation $U_f$ on $f$ and orthonormal ancillary states $|\chi_{\alpha,j}\rangle_{f_2^\alpha\bar{f}}\in \mathcal{H}_{f_2^{\alpha}\bar{f}}$ such that $U_f|\widetilde{\alpha,ij}\rangle = |\alpha,i\rangle_{f_1^\alpha}\otimes |\chi_{\alpha,j}\rangle_{f_2^\alpha\bar{f}}$.
\end{enumerate}
\end{theorem}
In other words, accessibility of logical operators on $f$ (statement 1) is equivalent to state-level recoverability from $f$ (statement 2). Therefore, to understand what information is retrievable on a subsystem $f$, it suffices to examine the accessible logical subalgebra $M_f$ supported on $f$. A similar statement is found in \cite{Riedel_2017}. 

\paragraph*{Classicality, Redundancy and Quantum Codes.}
The accessible subalgebra $M_f$ allows us to classify objectivity through information classicality and redundancy more precisely with known structures from quantum coding theory.

\begin{definition}
\label{def:classical_info}
    For a fixed partitioning of physical Hilbert space $\mathcal{H}_P=\bigotimes_{i=1}^N\mathcal{H}_{f_i}$, we say that a fragment $f_i$ contains classically retrievable logical information if the center of its recoverable logical algebra $Z(M_{f_i})$ is nontrivial, i.e., contains elements not proportional to the logical identity.
\end{definition}
Logical operators in the center are accessible on disjoint fragments $f_i$ and $\bar{f_i}$; hence measuring the record on $f_i$ does not disturb the measurement on its complement. Since the center is abelian, the joint eigenspaces of $Z(M_{f_i})$ define the classical sectors of the recoverable algebra, and projective measurement of any generator in the center on $f_i$ recovers the corresponding sector label $\alpha$ without disturbing any other observable in $M_{f_i}$.
This is in contrast with the quantum information contained in $f_i$, which is sensitive to the order of measurements, and by the no-cloning theorem, grants $f_i$ exclusive access. 

In the repetition code example, $\tilde{Z}\in Z(M_{f_i})$ is equivalent to the recoverability of a classical state $|a|^2|0\rangle\langle 0|+|b|^2|1\rangle\langle 1|$ in equation~\ref{eqn:ghzrecov} by Theorem~\ref{thm:harlow}.  On the other hand, the quantum information of the system is retrievable only when the full logical algebra is accessible, thus requiring one to have access to the anti-commuting pair $\tilde{X}$ and $\tilde{Z}$.

This algebraic view of classical versus quantum can be made concrete on any stabilizer code.
\begin{theorem}
    \label{thm:stabilizer_algebra}
   Let $\mathcal{S}\subset\mathcal{P}^n$ be an abelian subgroup of the $n$-qubit Pauli group that defines a stabilizer code, and $M_f$ is the logical Pauli subalgebra with support on  $f$. Then $M_f$ can always  be generated by logical Pauli operators such that $M_f= \langle \tilde P_1, \tilde Q_1,\tilde P_2,\tilde Q_2,\dots, \tilde P_q, \tilde Q_q, \tilde T_1,\tilde T_2,\dots, \tilde T_c\rangle$ where $\{\tilde P_i,\tilde Q_i\}=0$ form $q$ anticommuting pairs and $\tilde T_j$ commute with all other generators. Furthermore, $M_f$ can access $c=\operatorname{rank}(Z(M_f))$ bits of encoded classical information and $q$ qubits of encoded quantum information.
\end{theorem}
We refer to $q$ as the \emph{quantum rank}.
The proofs of all theorems can be found in the Supplemental Materials \cite{supplemental}. To connect with existing information-theoretic measures used to define classical plateaus, we also examine its property in the state picture. 

Let $V_{\mathcal S}:\mathcal{H}_S\rightarrow \mathcal{H}_{SE}$ be the encoding isometry of the stabilizer code and let $|\Phi^+\rangle_{RS}=2^{-k/2}\sum_{i}|ii\rangle_{RS}$ be the maximally entangled state between $\mathcal{H}_S$ and an isomorphic reference $\mathcal{H}_R$, then the Choi state $|V_{\mathcal{S}}\rangle_{RSE} = V_{\mathcal{S}} |\Phi^+\rangle_{RS}$  of $V_\mathcal{S}$  satisfies the following theorem for any $f\subseteq SE$.
\begin{theorem}
    The QMI between the reference $R$ and fragment $f$ is 
    \begin{align}
        I(R:f) =\operatorname{rank}(M_f)=2q+c.
    \end{align}
    \label{thm:stab_QMI}
\end{theorem}
The contributions to the QMI in stabilizer states come from two origins where each quantum bit contributes two units of QMI and each classical bit contributes one. 

Since the classicality of information comes from commuting logical operators in the center whereas quantum information comes from anticommuting pairs, one can easily identify codes where small fragments have subalgebras which do not contain any anticommuting pairs. The prime examples are asymmetric Calderbank-Shor-Steane (CSS) codes  \cite{asym_code1,asym_code2} where, without loss of generality, small fragments have access to only $Z$-type logical operators whereas a much larger fragment is needed to access the $X$-type operators required to complete the anticommuting pairs. The quantum repetition code is a special instance of such codes. The degree of asymmetry can be understood as an order parameter that separates the QD phase from the quantum encoding phase in \cite{ferte_quantum_2024}.

\begin{theorem}
\label{theorem:distance_css_prop}
    An asymmetric $[[n,1,d_X,d_Z]]$ CSS code with distances $d_X> d_Z$ has a classical plateau measured by $\max_fI(R:f)$ with width $w\geq |d_X-d_Z|$.
\end{theorem}
Both the generalized Shor code and the rectangular surface code \cite{Bacon_2006,Bacon2,Fowler_2012} satisfy the above condition as long as we choose the width and height to be different. 
This follows directly from the definition of the asymmetric distances: for \(d_Z\le |f|<d_X\), any such fragment can support \(\widetilde Z\) but not \(\widetilde X\), so Theorem~\ref{thm:stab_QMI} gives \(I_{\max}(R:f)=1\).

More systematically, consider a random construction using any $[n,k,d]$ classical linear code  with generator matrix $G\in\mathbb{F}_2^{k\times n}$ and Hamming distances $d$. Define the generating matrix as
\begin{align}
    G = [I_k|P], \qquad P \in \mathbb{F}_2^{k\times (n-k)}.
    \label{eq:G_mat}
\end{align}
where $P$ can be chosen to be any binary matrix. Then, define the $Z$-check matrix as,
\begin{align}
    H_Z = [P^T|I_{n-k}],
    \label{eqn:randomcode}
\end{align}
and take $H_X$ to be empty. The matrix $H=H_Z\oplus H_X$ defines a CSS code whose stabilizer group is generated by only $Z$-type Pauli strings, $\mathcal{S}=\langle S_i=Z^{\mathbf{r}_i}\rangle$, where the binary vector $\mathbf{r}_i$ denotes the $i$-th row of $H_Z$. For a canonical choice of logical basis, it has $d_Z^{(i)}=1$ for all logical qubits $i=1,\dots k$ and $d_X^{(i)}\geq d$. Hence all such codes are good candidates for models of objectivity. An example with $k>1$ is shown in Fig.~\ref{fig:QDQEC_plateaus}a. 

Alternative constructions with translation invariance using classical cyclic codes can also be constructed to support a classical plateau \cite{prange1957cyclic,long_paper} for all contiguous fragments.

The notion of classicality above detects whether information is copyable, but not whether many disjoint fragments can recover it. To recover objectivity, we also need the information to be \emph{redundant}.

\begin{definition}
\label{def:ell_redundancy}
    A classical bit is $\ell$-redundant if $\tilde T_i\in \bigcap_{j=1}^\ell Z(M_{f_j})$ is in the intersection of $\ell$ disjoint fragments from a non-trivial partitioning $\{f_j\}_{j=1}^{\geq \ell}$ of $SE$. 
\end{definition}
The relevant partitioning depends on the physical setup, such as spatial locality \footnote{We note that this is slightly different from the Zurekian definition where redundancy means the number of independent fragments of the environment that have access to almost all of the classical information.}.

Note that this redundancy condition is simply the local recoverability from $f_k$ of a code \cite{Riedel_2017} (with respect to a code subalgebra generated by $T_i$). Redundancy can then be quantified by the coset quantum weight enumerator polynomials.

\begin{definition}{\textbf{Coset enumerator: }}
    Let $P\in \mathcal P^n$ be a Pauli operator. The coset enumerator polynomial of a stabilizer code is 
\begin{equation}
    A_P(z) = \frac{1}{K^2}\sum_{Q\in \mathcal{P}^n}\Tr[Q\Pi_P]\Tr[ Q^\dagger\Pi_P]z^{wt(Q)}=\sum_{j=0}^n A_j^{(P)} z^j
\end{equation}
where $$\Pi = \frac{1}{|\mathcal{S}|}\sum_{s\in \mathcal S} s,~\Pi_P=P\Pi,~K=\Tr[\Pi]$$ and $wt(Q)$ is the weight (or operator size) of $Q$. 
\end{definition}

For $P=\tilde{Z}$ or $\tilde{X}$, the sequence $\{A_j^{(P)}\}$ records how many representatives of the corresponding logical operator exist at each weight \cite{cao_quantum_2023}. For an $[[n,1]]$ CSS code, Theorem~\ref{thm:stab_QMI} implies that the average QMI between $R$ and a random fragment $f$ is
\begin{align}
    \mathbb{E}_{|f|}[I(R:f)]=P_Z(|f|)+P_X(|f|),
\end{align}
where $P_{Z/X}(|f|)$ is the probability that $f$ contains the support of at least one representative of $\tilde Z/\tilde X$. 
\begin{theorem}[Averaged QMI]
\label{thm:avg_QMI}
  Assuming the supports of logical operators are uniformly distributed across the physical qubits, the average QMI from an $[[n,1]]$ CSS code is
  \begin{align}
    \mathbb{E}_{|f|}[I(R:f)]=2-\prod_{j=1}^{|f|}\frac{\binom{\binom{n}{j}-\binom{|f|}{j}}{A_j^{(\tilde{Z})}}}{\binom{\binom{n}{j}}{A_j^{(\tilde{Z})}}}-\prod_{j=1}^{|f|}\frac{\binom{\binom{n}{j}-\binom{|f|}{j}}{A_j^{(\tilde{X})}}}{\binom{\binom{n}{j}}{A_j^{(\tilde{X})}}}.
    \label{eqn:cosetavgQMI}
\end{align}
Furthermore, if  $A_j^{(\tilde{Z})},A_j^{(\tilde{X})}\ll\binom{n}{j}$, then 
\begin{align}
    \mathbb{E}_{|f|}[I(R:f)]\approx(1-e^{-\lambda_Z(|f|)})+(1-e^{-\lambda_X(|f|)}),
    \label{eq:avg_QMI_exact}
\end{align}
with
\begin{align}
    \lambda_Z(|f|)=\sum_{j\leq |f|} A_j^{(\tilde{Z})} \frac{\binom{|f|}{j}}{\binom{n}{j}}, \qquad
    \lambda_X(|f|)=\sum_{j\leq |f|} A_j^{(\tilde{X})} \frac{\binom{|f|}{j}}{\binom{n}{j}}.
    \label{eq:lambda}
\end{align}

\end{theorem}

The functions $\lambda_{Z/X}(|f|)$ quantify the probability with which a random fragment contains a representative of the corresponding logical operator, assuming the chance of finding a logical operator is uniform across all fragments of the same size. This statement is robust to small deviations from uniformity (See SM).  We verify the above numerically (Fig.~\ref{fig:QDQEC_plateaus}b) where the assumption is valid for the typical asymmetric codes we construct. A plateau appears when the $Z$-coset has become typical and the $X$-coset has not, e.g., when $\lambda_Z\gg \lambda_X$. We present a generalized argument in Ref.~\cite{long_paper} for $[[n,k>1]]$ CSS codes.

\begin{corollary}
\label{cor:avg_plateau}
For any $0<\epsilon<\frac{1}{2}$, suppose
    \begin{align}
        m_Z^{(\epsilon)}
        &:=
        \min\{m\in \mathbb{Z}:\lambda_Z(m)\geq \log(1/\epsilon)\},\\
        m_X^{(\epsilon)}
        &:=
        \min\{m\in\mathbb Z:\lambda_X(m)\geq-\log(1-\epsilon)\},
    \end{align}
    and $m_Z^{(\epsilon)}<m_X^{(\epsilon)}$. Then the average QMI has a plateau of width $m_X^{(\epsilon)}-m_Z^{(\epsilon)}$ with vertical range $[1- \epsilon,1+\epsilon]$. 
\end{corollary}
Fig.~\ref{fig:QDQEC_plateaus} shows the bounds proposed by Corollary~\ref{cor:avg_plateau} on a plateau with $\epsilon=0.1$.

\paragraph*{Algebraic objectivity.}
In addition to their strong connection to existing QD concepts, the algebraic formulation provides precise tracking of both the type and the distribution of information --- given $f$, one can efficiently determine the logical operators supported on it using our algorithm presented in the End Matter. With this improved precision, we can easily classify different strengths of objectivity and characterize them along our two criteria: classicality and redundancy. We highlight a few interesting classes and draw connections with existing literature on QD.

\begin{figure}
  \centering
\resizebox{1.0\linewidth}{!}{\input{qd_diagram_v4.tikz}}
 \caption{Algebraic objectivity regimes for a fragment family $\mathcal F=\{f_i\}$. The horizontal axis counts $\ell$-redundant classical bits with $\ell=O(|\mathcal{F}|)$, while the vertical axis gives the typical local quantum rank $\bar q=\langle q_i\rangle_{\mathcal F}$. Color encodes redundancy and quantum rank. Dashed lines indicate schematic regime boundaries; the hatched region is forbidden, as entering it would require the fragment family to redundantly share noncentral logical information.}
 \label{fig:obj_regimes}
\end{figure}
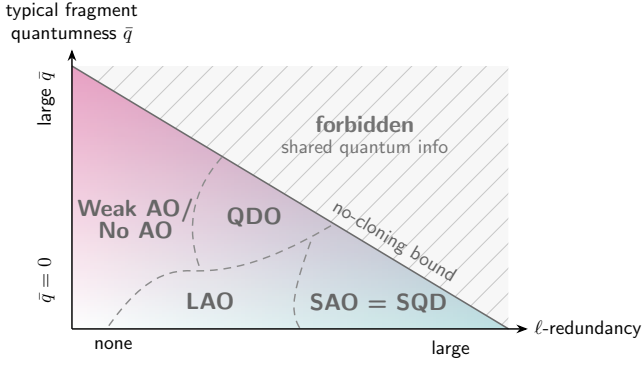

\begin{figure}
    \centering
    \includegraphics[width=1\linewidth]{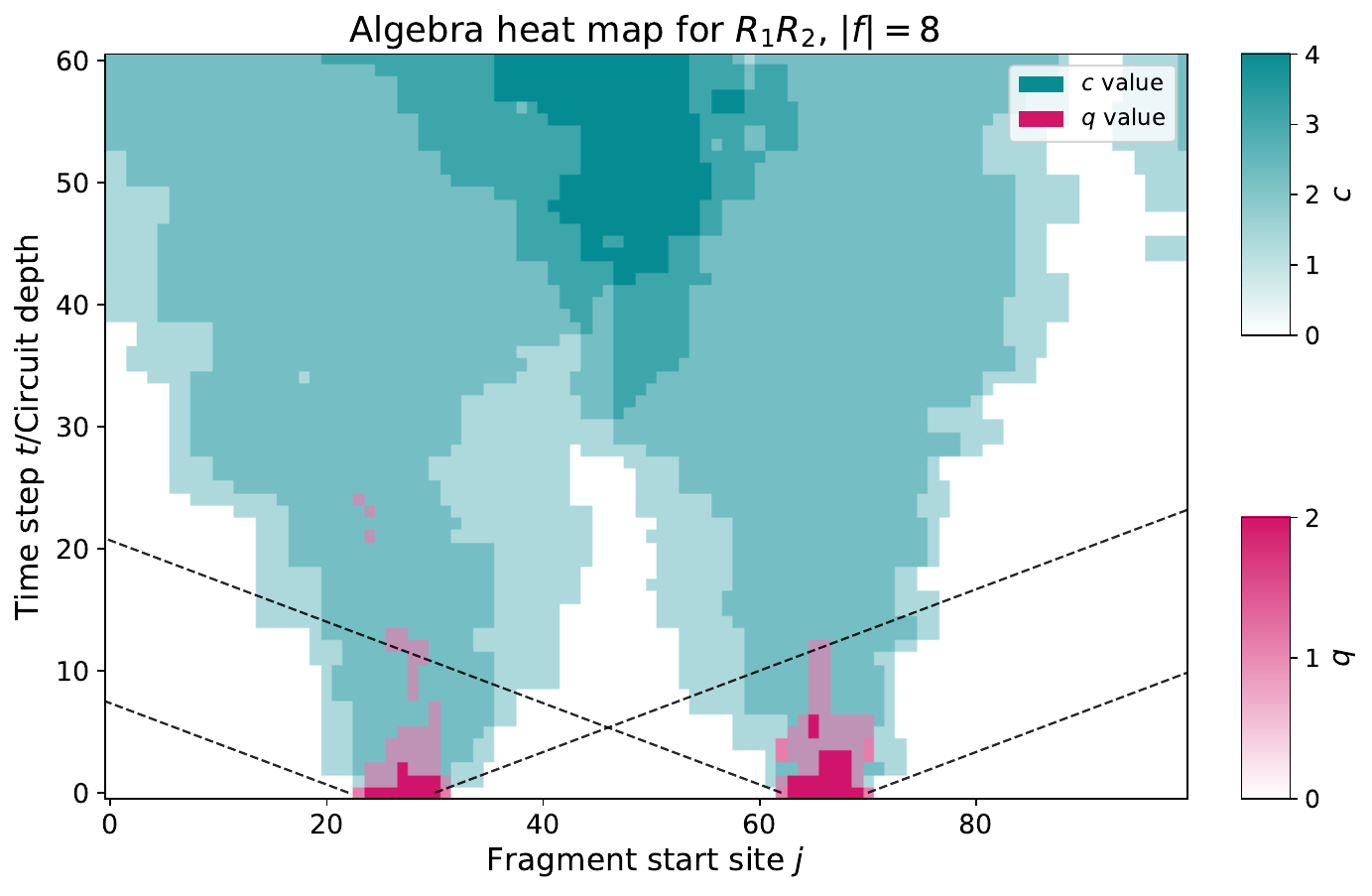}
    \caption{Heat map of the algebra $M_f$ for all contiguous fragments $f_j=\{j-|f|+1,j-|f|+2,\cdots,j\}$ with $|f|=8$ and two $k=2$ sources broadcasting in a brickwork circuit construction. Each pixel encodes the algebraic content with teal for classical $c$ information and pink for quantum information $q$. Classical records spread within the emergent light cones (dashed line), while quantum information is rapidly lost.}
    \label{fig:obj_cones}
\end{figure}

\begin{enumerate}[leftmargin=*]
    \item Strong algebraic objectivity (SAO) requires that all relevant fragments $\mathcal F=\{f_i\}$ have a common center that is maximal. 
\begin{equation}
    \bigcap_iZ(M_{f_i})=M_{f_j},~\forall j.
    \label{eqn:SAO}
\end{equation}
In other words, only classical information is found in each fragment, and the information is identical across all fragments in $\mathcal F$. 

This coincides with the original Zwolak-Zurek construction involving the GHZ state \cite{ZwolakZurek} where all fragments have the same information.

\item Localized algebraic objectivity (LAO), where all proliferated information is classical, i.e. $M_{f_i}=Z(M_{f_i})$ for all $i$, but the common center need not be maximal across different fragments: $M_{f_i}\cap M_{f_j}\subset Z(M_{f_i})$. This means that, for some partitioning, each fragment only contains classical information of the system, but different fragments know about different parts of the decohered system. 

This occurs, for example, in the random construction~(equation \ref{eqn:randomcode}) with $k>1$. A physically relevant example is when spatially separated systems decohere at different times. There is a limited notion of redundancy within each lightcone, but there is no shared information before the lightcones overlap. See Fig.~\ref{fig:obj_cones}. 
    \item Quantum doped objectivity (QDO), where we allow quantum information to be found in fragments such that $Z(M_{f_i})\subset M_{f_i}$. The quantum information is exclusive to each fragment, but the classical information can remain redundant either globally or in a localized fashion like in LAO. 

    Systems showing QDO correspond to fragments that may contain some anticommuting pairs in the supported logical subalgebra but remain dominated by classical information. For example, a quantum computer can operate within this regime where most of the underlying hardware substrate like superconducting circuit boards are classical objects, but can nevertheless sustain coherent qubits used for computation.
\end{enumerate}
By tuning the classicality of information and redundancy, one can arrive at different regimes of objectivity (Fig.~\ref{fig:obj_regimes}).

In certain regimes, the algebraic definition coincides with the familiar information-theoretic definitions:
\begin{theorem}
    Strong quantum Darwinism is equivalent to strong algebraic objectivity.
\end{theorem}

Both LAO and QDO can still retain a high degree of objectivity according to the definition by Horodecki, Korbicz, and Horodecki \cite{horodecki_quantum_2015}. They can overlap with, but do not always fit naturally within the conventional QD framework, because $S(R)\ne I(R:f)$; where a system can be constructed to violate the equality substantially.

For $k>1$, there exist other subtle regimes of objectivity and classicality that are invisible to a traditional QMI metric but merit a more in-depth discussion. We refer interested readers to \cite{long_paper} for details. 

\paragraph*{Clifford dynamics and objectivity cone.}
Another advantage of the QD-QECC correspondence is dynamical simulation --- by the Gottesman-Knill theorem \cite{gottesman_stabilizer_1997}, large and complex decoherence dynamics can now be recast as stabilizer simulations. Thus far, we have focused on using $U_{\rm SE}$ as the long time limit of system-environment interactions. However, more realistic dynamics are continuous and are local. 

By stacking the individual encoding unitaries of small quantum codes into a brickwork circuit (see End Matter), the underlying circuit dynamics now explicitly shows the proliferation of classical information bounded by the emergent light cone where conventional considerations of Lieb-Robinson velocity apply (Fig.~\ref{fig:obj_cones}). In this model, classical information propagates much slower than the maximum propagation speed allowed by the emergent lightcones (dashed lines). We see that quantum information quickly decoheres, but the classical information continues to proliferate throughout the simulated dynamics.

\paragraph*{Conclusion.}

In this letter, we established a novel connection between quantum foundations and QECC. Conceptually, we constructed a unifying algebraic framework for objectivity that removes previous ambiguities on classicality and redundancy. The precise tracking of quantum and classical information further enables new quantitative classification and analysis of emergent objectivity. 

On the technical level, known results on QECCs and the stabilizer formalism provide an efficient and powerful machinery for simulating and analyzing decoherence and the emergence of classicality through quantum dynamics. Together, these findings enable new coding-theoretic classification of decoherence processes and can support large-scale simulations.

Although we focus on qubit stabilizer codes for much of the paper, parts of the framework are extendable to more general operator algebras~\cite{long_paper}. Future work will explore models with $k>1$, where the information structure becomes richer, as well as extensions beyond Pauli stabilizer codes.  

\section*{End Matter}
\paragraph*{Dynamical model and algebra recovery.}

The dynamical data in Fig.~\ref{fig:obj_cones}, as well as more general decoherence dynamics via Clifford circuits, can be obtained from a stabilizer simulation. 

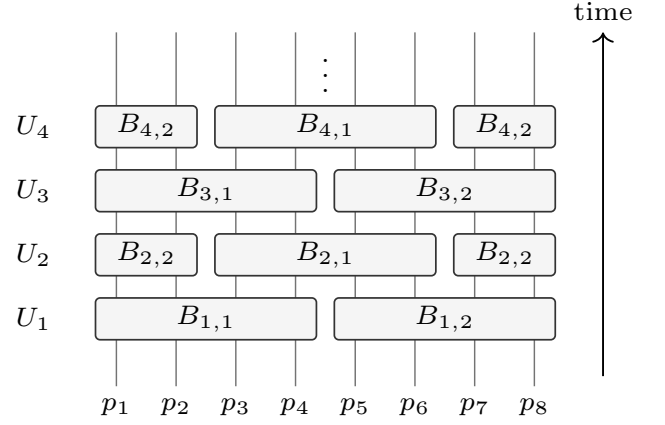
\begin{figure}[H]
    \centering
    \resizebox{0.98\columnwidth}{!}{%
    \begin{tikzpicture}[
        x=0.55cm,
        y=0.62cm,
        wire/.style={line width=0.38pt, draw=black!55},
        brick/.style={draw=black!80, line width=0.45pt, fill=black!4, rounded corners=1.2pt},
        layerlabel/.style={font=\scriptsize, anchor=east},
        bricklabel/.style={font=\scriptsize}
    ]

    \def\ytop{5.15}
    \def\h{0.31}
    \def\m{0.10}

    \foreach \x in {0,...,7} {
        \draw[wire] (\x,-0.10) -- (\x,\ytop);
        \pgfmathtruncatemacro{\q}{\x+1}
        \node[below,font=\scriptsize] at (\x,-0.10) {$p_{\q}$};
    }

    \node[layerlabel] at (-0.88,0.90) {$U_1$};
    \node[layerlabel] at (-0.88,1.85) {$U_2$};
    \node[layerlabel] at (-0.88,2.80) {$U_3$};
    \node[layerlabel] at (-0.88,3.75) {$U_4$};

    \draw[brick] (-0.45+\m,0.90-\h) rectangle (3.45-\m,0.90+\h);
    \node[bricklabel] at (1.50,0.90) {$B_{1,1}$};
    \draw[brick] (3.55+\m,0.90-\h) rectangle (7.45-\m,0.90+\h);
    \node[bricklabel] at (5.50,0.90) {$B_{1,2}$};

    \draw[brick] (1.55+\m,1.85-\h) rectangle (5.45-\m,1.85+\h);
    \node[bricklabel] at (3.50,1.85) {$B_{2,1}$};
    \draw[brick] (-0.45+\m,1.85-\h) rectangle (1.45-\m,1.85+\h);
    \node[bricklabel] at (0.50,1.85) {$B_{2,2}$};
    \draw[brick] (5.55+\m,1.85-\h) rectangle (7.45-\m,1.85+\h);
    \node[bricklabel] at (6.50,1.85) {$B_{2,2}$};

    \draw[brick] (-0.45+\m,2.80-\h) rectangle (3.45-\m,2.80+\h);
    \node[bricklabel] at (1.50,2.80) {$B_{3,1}$};
    \draw[brick] (3.55+\m,2.80-\h) rectangle (7.45-\m,2.80+\h);
    \node[bricklabel] at (5.50,2.80) {$B_{3,2}$};

    \draw[brick] (1.55+\m,3.75-\h) rectangle (5.45-\m,3.75+\h);
    \node[bricklabel] at (3.50,3.75) {$B_{4,1}$};
    \draw[brick] (-0.45+\m,3.75-\h) rectangle (1.45-\m,3.75+\h);
    \node[bricklabel] at (0.50,3.75) {$B_{4,2}$};
    \draw[brick] (5.55+\m,3.75-\h) rectangle (7.45-\m,3.75+\h);
    \node[bricklabel] at (6.50,3.75) {$B_{4,2}$};

    \node[font=\scriptsize] at (3.50,4.68) {$\vdots$};
    \draw[->, line width=0.5pt] (8.15,0.05) -- (8.15,\ytop);
    \node[above,font=\scriptsize] at (8.15,\ytop) {time};

    \end{tikzpicture}%
    }

    \caption{Schematic brickwork Clifford circuit with periodic boundary conditions. Each layer is a product of disjoint local bricks, and adjacent layers are shifted by half a block. Matching labels at the two edges denote a periodic brick.}
    \label{fig:brickwork}
\end{figure}

The general methodology is simple ---  we place $n$ physical qubits on a one-dimensional ring and choose $k$ of them as the system logical degrees of freedom. The remaining qubits are initialized to $|0\rangle$. Then, a brickwork circuit consisting of spatially local Clifford unitaries is applied to the $n$ qubits (Fig.~\ref{fig:brickwork}). For simplicity, we have chosen to implement circuits local in one spatial dimension, but by the Gottesman-Knill and the Aaronson-Gottesman construction \cite{gottesman_stabilizer_1997,Aaronson_2004}, such simulations remain efficient at higher spatial dimensions.

Let $U_l=\prod_\beta B_{l,\beta}$ be the unitary gates on the $l$-th layer of the circuit where each brick $B_{l,\beta}$ is a Clifford unitary acting on $m$ neighboring physical qubits. Within a layer, the bricks are disjoint, while consecutive layers are shifted by $m/2$ sites, as represented in Fig.~\ref{fig:brickwork}.

The encoding unitary for up to time $t$ is
\begin{align}
    \mathcal U_t=U_tU_{t-1}\cdots, U_1.
\end{align}

This dynamical circuit can be understood as a sequence of encoding unitaries $\mathcal U_1, \mathcal U_2,\dots \mathcal U_t$ one for each time $t$. 
For each $\mathcal U_t$, we have an associated encoding isometry $V_t: \mathcal{H}_S\rightarrow \mathcal{H}_{SE}$ where we simply fix the $n-k$ initial inputs of $\mathcal{U}_t$ to the $|0\rangle$ state.

This constitutes a local circuit model for objectivity: logical operators can spread only inside the brickwork light cone, while the algebraic decoder below determines how much total information of the system is contained inside a fragment, and how much of it is classical or quantum.

For the simulations shown in Fig.~\ref{fig:obj_cones}, each brick is chosen to be the encoding unitary of a small asymmetric CSS code, e.g., that of equation \ref{eqn:randomcode}. 
From the parity check matrix, one can determine the Clifford encoding unitary from \cite{gottesman_stabilizer_1997}.

\paragraph*{Algebraic recovery from a fragment.} To determine how much quantum or classical information is recoverable on a fragment $f$ at each time $t$, we determine which set of logical Pauli operators have representatives supported entirely inside of $f$. This can also be done efficiently through binary matrix operations.

For this task, it is conceptually simpler to use the Choi state $|V_t\rangle$ of the encoding map $V_t$ where
\begin{align}
    |\Phi^+\rangle_{RS}&=2^{-k/2}\sum_{i}|ii\rangle_{RS}\\
    |V_t\rangle_{RSE}&=(I_R\otimes V_t)|\Phi^+\rangle_{RS}.
\end{align}

Then a representation of logical $P$ is recoverable in $f$ if and only if $P_R\otimes \tilde{P}_{f}\otimes I_{\bar{f}}|V_t\rangle = |V_t\rangle$. In other words, the stabilizer element is supported entirely on $R\cup f$. See e.g. Proposition 2.3 of \cite{Cao:2026uwb}. Hence our task of finding local operators supported on $f$ in the code is reduced to looking for stabilizer elements with support on $R\cup f$ in the Choi state.

Let $\mathcal{S}_t$ be the (full rank) Pauli stabilizer group of $|V_t\rangle$ with $n+k$ generators $\{g_i = \omega(\mathbf{a}_i,\mathbf{b}_i)Z^{\mathbf{a}_i}X^{\mathbf{b}_i}\}$. Here $\mathbf{a},\mathbf{b}$ are length-$(n+k)$ binary strings such that $Z^{\mathbf{a}}X^{\mathbf{b}}$ is a Pauli string of length $n+k$ with the $j$th position taking $Z^{a_j}X^{b_j}$ and $j=1,\dots, n+k$. The complex phase $\omega$ is some power of $i$ which depends on the values of $\mathbf{a},\mathbf{b}$. As the phase information is not relevant for our algorithm, one can then represent $\mathcal{S}_t$ as a binary symplectic matrix $H_t$ where the $i$th row is $(\mathbf{a}_i|\mathbf{b}_i)$. The generators, i.e., check matrix, of $\mathcal{S}_t$ can be determined efficiently from the well-established Aaronson-Gottesman method \cite{Aaronson_2004}, which is also available in most quantum simulation packages \cite{Gidney_2021}.

For each $t$, the logical operator search can be done through a three-step algorithm:

\begin{enumerate}
    \item We first shuffle the columns of $H_t$ such that the columns (for both the Z and X sections) are arranged as $[\bar {f}|f|R]$, then perform Gaussian elimination over $\mathbb{F}_2$ to obtain its reduced row echelon form (RREF). Keeping only the rows that have no support in the columns corresponding to qubits in $\bar{f}$ and removing the zero columns, we obtain a reduced $H'_t=[f'|R']$. This keeps only generators of logical operators and stabilizers with support in $f$.
    \item Next, we arrange the columns of $H'_t$ with order $[R'|f']$, then perform row elimination again to get its RREF. Keeping only rows with support in $R'$ and removing zero columns, we obtain a check matrix $H''_t$ with the form $[R''|f'']$ for both the Z and X sections. This then eliminates any stabilizer generators that don't act non-trivially on the logical space. 
    \item Finally, the rows of $[f'']$ correspond to a logical basis with support only in $f$. Performing the standard row operation of \cite{cao_quantum_2025}, one can arrange it to the standard form with $2q$ rows that correspond to the $q$ symplectic pairs and $c$ rows that correspond to elements in the center.

\end{enumerate}
By row counting, we can then determine precisely the qubits and bits recoverable from $f$. This can be thought of as a simplification of the Gaussian elimination decoder \cite{Steinberg_2025} using Choi states.

The most expensive step of the above algorithm is Gaussian elimination, which has complexity no more than $O((n+k)^3)$ for each time and fragment. The overall generation of such information dynamics is approximately $O(TN_f(n+k)^3)$ where $T$ is the total number of time steps and $N_f$ is the number of fragments we check per time step.

\section*{Acknowledgement}
We thank Sean M. Carroll, Aidan Chatwin-Davies,  Giuseppe Cotardo, Jordan Cotler, and Jason Pollack for helpful comments and discussions. M.G. acknowledges support from the Commonwealth Cyber Initiative Southwest Virginia.

\nocite{bravyi_ghz_2006}
\bibliography{QEC_QD}
\bibliographystyle{apsrev4-2}

\section{Supplemental material}

\subsection{Proof of Theorem 3}
We represent logical Pauli operators modulo phases and stabilizers by binary vectors, with the usual symplectic form encoding commutation, adopting the definition and results from [32].%\cite{cao_quantum_2025}. 

Let $W_f$ be the binary symplectic vector subspace corresponding to the logical Pauli operators in $M_f$, namely those admitting representatives supported on $f$. Then the center of $M_f$ is represented by
\begin{align}
    \operatorname{rad}(W_f)=W_f\cap W_f^\perp ,
\end{align}
because these are exactly the elements of $W_f$ that commute with all elements of $W_f$.
By the standard symplectic normal form, any subspace $W_f$ admits an orthogonal decomposition
\begin{align}
    W_f=\operatorname{rad}(W_f)\oplus K_f ,
\end{align}
where $K_f$ is symplectic [Thm.~1.6 in 32]. Choosing a
symplectic basis $p_1,q_1,\ldots,p_q,q_q$ for $K_f$ and a basis $t_1,\ldots,t_c$ for $\operatorname{rad}(W_f)$, and lifting these vectors back to logical Pauli operators, gives
\begin{align}
    M_f=\left\langle \widetilde P_1,\widetilde Q_1,\ldots,\widetilde P_q,\widetilde Q_q, \widetilde T_1,\ldots,\widetilde T_c \right\rangle ,
\end{align}
with $\{\widetilde P_i,\widetilde Q_i\}=0$ and all remaining commutators trivial. The central generators are
\begin{align}
    Z(M_f)=\langle \widetilde T_1,\ldots,\widetilde T_c\rangle .    
\end{align}
Hence $M_f$ decomposes into $q$ anti-commuting logical Pauli pairs, allowing one to manipulate the information on $q$ encoded qubits. The $c$ independent commuting logical Paulis allow one to perform only a phase operation in some chosen basis, thus accessing $c$ encoded classical bits. To see it from the state recovery perspective, we simply apply Theorem 1. 

\subsection{Proof of Theorem 4}
Let $\mathcal{S}$ be the stabilizer group of the code, and let
\begin{align}
    |V_{\mathcal S}\rangle_{RSE}=(\mathbb I_R\otimes V_{\mathcal S})|\Phi^+\rangle_{RS}
\end{align}
be the Choi state of the encoding isometry, with

\begin{align}
    |\Phi^+\rangle_{RS}=2^{-k/2}\sum_{i=1}^{2^k}|ii\rangle_{RS}.
\end{align}
$|V_{\mathcal S}\rangle_{RSE}$ is a stabilizer state on $RSE$.

Denote its stabilizer group by $\mathcal T$. For any subsystem $A\subseteq RSE$, let
\begin{align}
    \mathcal T_A:=\{T\in \mathcal T:\operatorname{supp}(T)\subseteq A\}
\end{align}
be the subgroup of Choi stabilizers supported entirely on $A$. We write $\dim\mathcal T_A$ for its binary rank. Using the standard entropy formula for stabilizer states [34]: for a pure stabilizer state and any subsystem $A$, its von Neumann entropy is
\begin{align}
    S(A)=|A|-\dim\mathcal T_A.
\end{align} 
Applying this formula to the tripartition $Rf\bar f$, we get
\begin{align}
    I(R:f)
    &= S(R)+S(f)-S(Rf)  \\
    &= \bigl(|R|-\dim\mathcal T_R\bigr)+\bigl(|f|-\dim\mathcal T_f\bigr)\\
    &-\bigl(|R|+|f|-\dim\mathcal T_{Rf}\bigr).
\end{align}
Since, by construction, the reference $R$ is maximally mixed in the Choi state, there are no nontrivial stabilizers supported only on $R$, so $\mathcal T_R=0$. Hence
\begin{align}
    I(R:f)=\dim\mathcal T_{Rf}-\dim\mathcal T_f .
\end{align}
It remains to identify the right-hand side with the rank of the logical
Pauli algebra supported on $f$. Consider the projection map from $\mathcal T_{Rf}$ to $\mathcal{P}_R$ the Pauli strings with support only on $R$
\begin{align}
    \pi_R:\mathcal T_{Rf}\longrightarrow \mathcal P_R,
\end{align}
which restricts a Choi stabilizer supported on $Rf$ to its reference component. Its kernel is precisely $\mathcal T_f$, since these are exactly the stabilizers in $\mathcal T_{Rf}$ acting trivially on $R$.
Therefore
\begin{align}
    \dim\operatorname{im}\pi_R=\dim\mathcal T_{Rf}-\dim\mathcal T_f .
\end{align}
We now show that $\operatorname{im}\pi_R$ is exactly the space generated by products of the generators of $M_f$. For a logical Pauli $P$, the maximally entangled state satisfies
\begin{align}
    (P_R\otimes I_S)|\Phi^+\rangle_{RS}=(I_R\otimes P_S^T)|\Phi^+\rangle_{RS},
\end{align}
and for Pauli operators the transpose only changes an irrelevant phase. 
After applying the encoding isometry,
\begin{align}
    (P_R\otimes I_{SE})|V_{\mathcal S}\rangle=(I_R\otimes \widetilde P_{SE})|V_{\mathcal S}\rangle ,
\end{align}
where $\widetilde P$ is the encoded physical representative of the logical Pauli $P$. Thus a reference Pauli $P_R$ is equivalent on the Choi state to the corresponding encoded logical Pauli. Consequently, $P$ admits a representative $P_f$ supported on $f$ if and only if
\begin{align}
    P_R P_f \in \mathcal T .
\end{align}
Equivalently, $P_RP_f$ is a Choi stabilizer supported on $Rf$. Hence the image of $\pi_R$ is precisely the space of logical Pauli operators that admit representatives supported on $f$, namely $M_f$. Therefore
\begin{align}
    I(R:f)=\dim\mathcal T_{Rf}-\dim\mathcal T_f=\operatorname{rank}(M_f).
\end{align}
By Theorem 3, its rank is
\begin{align}
    \operatorname{rank}(M_f)=2q+c,
\end{align}
meaning
\begin{align}
    I(R:f)=\operatorname{rank}(M_f)=2q+c.
\end{align}

\subsection{Proof of Theorem 5}
The asymmetric distances $d_Z, d_X$ in the CSS code indicate that there must exist at least one representative of logical $Z$ (resp. logical $X$) operator whose minimum weight is $d_Z$ (resp. $d_X$). Therefore, for any subset of qubits $A$ with $|A|<\min\{d_Z,d_X\}$, it cannot contain any logical operator. Hence the QMI must be zero by Theorem 4. For any $A$ with $d_Z\leq |A|<d_X$, it can contain at most the support of the logical $Z$ operator, and there must exist at least one subset $A$ that contains the support of logical $Z$, hence yielding $\max_f I(R:f)=1$. For $A$ where $|A|\geq d_X$, there must be at least one fragment that contains the logical $X$. However, it is not guaranteed that the fragment will also contain the representation of logical $Z$. An explicit example of this is the surface code, where the minimum support of logical $X$ intersects, but does not contain, the logical $Z$ operator. Therefore, the max QMI is at least $1$ but will not jump to $2$ until the same fragment is guaranteed to contain both logical $X$
and $Z$. Hence the max-QMI plateau has width lower bounded by $|d_X-d_Z|$.

\subsection{Proof of Theorem 8}
\label{ap:enssemble_avg_proof}

Let $f$ be a uniformly random fragment of fixed size $|f|$. For an $[[n,1]]$ stabilizer code with a particular representation of the logical operators $\tilde{X},\tilde{Z}$, we define the event $E_Z(f)$ to be the event that the coset $\tilde{Z}\mathcal{S}$ contains at least one representative supported in $f$, and define $E_X(f)$ similarly. In this setting,
\begin{align}
    I(R:f)&=\textbf{1}_{E_Z(f)}+\textbf{1}_{E_X(f)}.
\end{align}
Therefore, averaging over all fragments of size $|f|$ yields
\begin{align}
    \mathbb{E}_{|f|}[I(R:f)]=P_Z(|f|)+P_X(|f|),
    \label{eqn:avg_qmi}
\end{align}
with $P_Z(|f|)=Pr(E_Z(|f|))$ being the probability for $f$ to contain a representative of logical $Z$ and similarly for $P_X(|f|)$. 

At a fixed weight $j$, the set of supports of representatives in $\tilde Z\mathcal S$ is modeled as a uniformly chosen subset of size $A_j^{(\tilde Z)}$ from the $\binom{n}{j}$ possible $j$-subsets of physical qubits. Among these, $\binom{|f|}{j}$ supports are contained in the fragment $f$. Therefore, the event that none of the weight-$j$ representatives is supported in $f$ is the event that all $A_j^{(\tilde Z)}$ chosen supports lie in the complementary set of size $\binom{n}{j}-\binom{|f|}{j}$. This gives the probability
\begin{align}
    \frac{\binom{\binom{n}{j}-\binom{|f|}{j}}{A_j^{(\tilde Z)}}}{\binom{\binom{n}{j}}{A_j^{(\tilde Z)}}}.
\end{align}
Assuming independence of supports across different weights, the probability that no logical ${Z}$ of any weight appears in $f$ is the product over all $j$, therefore
\begin{align}
    P_Z(|f|)\approx 1-\prod_{j=1}^{|f|}\frac{\binom{\binom{n}{j}-\binom{|f|}{j}}{A_j^{(\tilde Z)}}}{\binom{\binom{n}{j}}{A_j^{(\tilde Z)}}},
\end{align}
with an analogous formula for $P_X(|f|)$. Substituting into Eq.~\ref{eqn:avg_qmi}, this gives 
\begin{align}
    \mathbb{E}_{|f|}[I(R:f)]=2-\prod_{j=1}^{|f|}\frac{\binom{\binom{n}{j}-\binom{|f|}{j}}{A_j^{(\tilde{Z})}}}{\binom{\binom{n}{j}}{A_j^{(\tilde{Z})}}}-\prod_{j=1}^{|f|}\frac{\binom{\binom{n}{j}-\binom{|f|}{j}}{A_j^{(\tilde{X})}}}{\binom{\binom{n}{j}}{A_j^{(\tilde{X})}}}
\end{align}
In the sparse regime $A_j^{(\tilde{Z})}\ll\binom{n}{j}$, sampling with or without replacement agrees to leading order and,
\begin{align}
    \frac{\binom{\binom{n}{j}-\binom{|f|}{j}}{A_j}}{\binom{\binom{n}{j}}{A_j}}\approx ((1-\frac{\binom{|f|}{j}}{\binom{n}{j}}))^{A_j}\approx \exp\big[-A_j^{(\tilde{Z})}\frac{\binom{|f|}{j}}{\binom{n}{j}}\big].
\end{align}
Thus,
\begin{align}
P_Z(|f|) \approx1-\exp[-\lambda_Z(|f|)],\quad \lambda_Z(|f|)=\sum_{j\leq |f|}A_j^{(\tilde Z)}\frac{\binom{|f|}{j}}{\binom{n}{j}},
\end{align}
and similar for $P_X(|f|)$. This proves Theorem 8.
\qed\\

\paragraph*{Robustness to deviations from uniformity.}
Uniformity enters the estimate via the probability that a randomly chosen fragment $f$ of size $|f|$ contains a representative support of a given weight $j$,
\begin{align}
    p_j(|f|)=\frac{\binom{|f|}{j}}{\binom{n}{j}}.
\end{align}
Small non-uniformity replaces this by $p_j(|f|)+\epsilon_j(|f|)$, thereby perturbing the effective rate
\begin{align}
    \lambda_Z(|f|)=\sum_{j\leq |f|}A_j^{(\tilde Z)}p_j(|f|),
\end{align}
by
\begin{align}
    \Delta\lambda_Z(|f|)=\sum_{j\leq |f|}A_j^{(\tilde Z)}\epsilon_j(|f|),
\end{align}
and analogously for $X$. Hence such perturbations merely shift the onset scale of $P_Z$ and $P_X$.

\subsection{Proof of Corollary 9}

In the sparse regime, Theorem 8 gives
\begin{align}
    \mathbb E_{|f|=m}[I(R:f)]\approx (1-e^{-\lambda_Z(|f|)})+(1-e^{-\lambda_X(|f|)}).
\end{align}
For \(m\ge m_Z^{(\epsilon)}\), the first term is at least \(1-\epsilon\), while for \(m<m_X^{(\epsilon)}\), the second term is at most \(\epsilon\). Hence, for every \(m_Z^{(\epsilon)}\le m<m_X^{(\epsilon)}\),
\begin{align}
    1-\epsilon\leq\mathbb E_{|f|=m}[I(R:f)]\leq 1+\epsilon .
\end{align}
The plateau width is therefore at least \(m_X^{(\epsilon)}-m_Z^{(\epsilon)}\).

\subsection{Proof of Theorem 10}
By Le's [5], the sQD condition can be defined for an ensemble of disjoint fragments $\mathcal{F}=\{f_j\}$ as the condition that for all fragments,
\begin{align}
    I(R:f_i)=I_{\mathrm{acc}}(R:f_i)=\chi(R:f_i)=S(R).
\end{align}
We will verify the equivalence between this condition and the condition of strong algebraic objectivity. 

Let $R$ be the reference system that we used in earlier proofs and let $\mathcal{F}=\{f_j\}$ be a family of disjoint fragments. For each fragment $f$, the recoverable logical algebra induces a Wedderburn decomposition
\begin{align}
    \mathcal{C}&\simeq \bigoplus_\alpha\mathcal{H}_{f^{\alpha}}\otimes \mathcal{H}_{\bar{f}^\alpha}\\
    M_{f}&\simeq\bigoplus_\alpha L(\mathcal{H}_{f^\alpha})\otimes I_{\bar{f}^\alpha}
\end{align}
whose center is generated by the sector label $\alpha$. Write
\begin{align}
    p_\alpha=\frac{d_\alpha \bar{d_{\alpha}}}{d_{\rm code}},\quad d_\alpha=\dim(\mathcal{H}_{f^\alpha}),\quad \bar{d_\alpha}=\dim(\mathcal{H}_{\bar{f}^\alpha}),
\end{align}
and define
\begin{align}
    C=H(\{p_\alpha\}),\quad Q=\sum_{\alpha}p_\alpha\log d_\alpha,\quad \bar{Q}=\sum_{\alpha}p_\alpha\log \bar{d_\alpha}.
\end{align}
For the Choi state of the encoding, the operator-algebra QEC entropy identity [13]] gives
\begin{align}
    I(R:f)=C+2Q,\quad S(R)=C+Q+\bar{Q}.
    \label{eqn:OAQEC_entropy_eq}
\end{align}
%S(R) is correct, and it implies I(R:f_j).
In the stabilizer setting, $C$ and $Q$ reduce to $c$ and $q$ counting central generators and anti-commuting generator pairs.\\

Assuming sQD for each fragment in $\mathcal{F}$,
\begin{align}
    I(R:f_j)=\chi(R:f_j)=S(R).
\end{align}
In a general OAQEC, the Holevo information can use both the classical sector label and distinguishable states inside the quantum sector $\mathcal{H}_{f_j^\alpha}$, but not the complementary factor. Hence
\begin{align}
    \chi(R:f_j)\leq C_j+Q_j,
\end{align}
as it is bounded by the entropy of the accessible degrees of freedom. 
We combine this bound with Eq.~\ref{eqn:OAQEC_entropy_eq} to get
\begin{align}
    C_j+2Q_j=I(R:f_j)=\chi(R:f_j)\leq C_j+Q_j.
\end{align}
Since $Q_j\geq 0$, we must have $Q_j=0$. Thus no non-commuting set of operators is reconstructable on the fragment. The equality $I(R:f_j)=S(R)$ then gives $C_j=S(R)$, and Eq.~\ref{eqn:OAQEC_entropy_eq} also gives $\bar{Q}_j=0$. Thus every sector is 1-dimensional on both sides, making the recoverable abelian algebra maximal. These maximal centers are necessarily the same by a no-cloning argument. Suppose $f_j$ and $f_i$ are two disjoint fragments with maximal centers $Z(M_{f_j})$ and $Z(M_{f_i})$. Then there exist $O_j\in Z(M_{f_j})$ and $O_i\in Z(M_{f_i})$ s.t $[O_i,O_j]\neq 0$. Yet $O_j$ and $O_i$ have representations with disjoint support, which necessarily commute. This is a contradiction, hence all fragments in $\mathcal{F}$ reconstruct one common maximal center, which is strong algebraic objectivity.\\

For the reverse direction --- assuming strong algebraic objectivity: for every $j$, $M_{f_j}=\bigcap_iZ(M_{f_i})$. Then the Wedderburn decomposition has only one-dimensional sectors, meaning $Q_j=\bar{Q}_j=0$ and $S(R)=C_j$. Eq.~(\ref{eqn:OAQEC_entropy_eq}) gives
\begin{align}
    I(R:f_j)=S(R).
\end{align}
Conditional fragment states associated with distinct central projectors are supported on orthogonal sectors. Measuring these projectors therefore distinguishes the sectors perfectly. The Holevo bound is saturated and
\begin{align}
    \chi(R:f_j)=I_{\rm acc}(R:f_j)=C_j=S(R).
\end{align}
Each fragment thus satisfies the conditions of strong quantum Darwinism.

We note that the equality involving the accessible information $I_{\rm acc}$ is not needed in an exact algebraic setting. It only becomes nontrivial for approximate records, noisy recovery, or approximately commuting operator algebras.

\end{document}

%% file: qd_diagram_v4.tikz
\begin{tikzpicture}[
    font=\sffamily\large,
    >=Stealth,
    axis/.style={line width=0.75pt, -{Stealth[length=2.1mm]}},
    bound/.style={line width=0.8pt, draw=black!58},
    phase/.style={line width=0.75pt, dashed, dash pattern=on 4pt off 3pt, draw=black!45},
    hatch/.style={draw=black!22, line width=0.28pt},
    region/.style={align=center, font=\sffamily\normalsize}
]

\definecolor{qdBlue}{RGB}{48,117,173}
\definecolor{qdTeal}{RGB}{0,112,116}
\definecolor{qdGreen}{RGB}{0,128,70}
\definecolor{qdOrange}{RGB}{201,97,18}
\definecolor{qdPink}{RGB}{206,36,116}
\definecolor{qdBlueish}{RGB}{6,140,147}

\def\W{7.9}
\def\H{4.75}

% Allowed region: shared classical center can coexist with local quantum
% degrees of freedom, up to the fixed logical budget.
\begin{scope}
    \clip (0,0) -- (0,\H) -- (\W,0) -- cycle;

    % Redundancy/common-center contribution: grows from left to right.
    \shade[
        left color=white,
        right color=qdBlueish!72
    ] (0,0) rectangle (\W,\H);

    % Quantumness contribution: grows from bottom to top.
    \shade[
        bottom color=white,
        top color=qdPink!72,
        opacity=0.62
    ] (0,0) rectangle (\W,\H);

    % Tiny wash only for label readability.
    \fill[white, opacity=0.04] (0,0) -- (0,\H) -- (\W,0) -- cycle;
\end{scope}
% Forbidden region: large common center and large typical quantum rank would
% require redundantly sharing noncentral logical operators.
\begin{scope}
    \clip (0,\H) -- (\W,\H) -- (\W,0) -- cycle;
    \fill[black!3] (0,\H) -- (\W,\H) -- (\W,0) -- cycle;

    \begin{scope}
        \clip (0,\H) -- (\W,\H) -- (\W,0) -- cycle;
        \foreach \x in {-4.0,-3.55,...,8.35}
            \draw[hatch] (\x,0) -- ++(5.6,5.6);
    \end{scope}
\end{scope}

% No-cloning/capacity boundary.
\draw[bound] (0,\H) -- (\W,0)
    node[pos=0.72, above, sloped, font=\sffamily\normalsize, text=black!62]
    {no-cloning bound};

% Curved dashed regime boundaries, clipped to the allowed region.
\begin{scope}
    \clip (0,0) -- (0,\H) -- (\W,0) -- cycle;

    % Classical/quantum separator.
    \draw[phase]
        (0.04,-1.00)
        .. controls (0.85,0.55) and (1.45,1.08) ..
        (2.15,1.05)
        .. controls (3.05,1.02) and (3.85,1.35) ..
        (5.00,2.05);

    % weak AO / QDO separator.
    \draw[phase]
        (2.30,1.14)
        .. controls (2.0,2) and (2.25,2.85) ..
        (3.22,3.55)
        .. controls (2.88,3.80) and (3.25,4.05) ..
        (3.18,4.25);

    % LAO / SAO separator.
    \draw[phase]
        (4.12,0.02)
        .. controls (3.92,0.34) and (4.02,0.75) ..
        (4.35,1.66);
\end{scope}

% Regime labels.
\node[region, text=black!60] at (3.30,1.85) {
    {\large\bfseries QDO}\\[-1pt]
    %{\normalsize $c_{\rm com}>0,\ \bar q>0$}
};

\node[region, text=black!60] at (2.50,0.32) {
    {\large\bfseries LAO}\\[-1pt]
    %{\normalsize local classical}
};

\node[region, text=black!60] at (5.55,0.30) {
    {\large\bfseries SAO = SQD}\\[-1pt]
};

\node[region, text=black!60] at (1.15,2.00) {
    {\large\bfseries Weak AO/}\\[-1pt]
    {\large\bfseries No AO}
};

\node[region, text=black!55] at (5.30,3.48) {
    {\large\bfseries forbidden}\\[-1pt]
    {\normalsize shared quantum info}
};

% Axes.
\draw[axis] (0,0) -- (8.25,0)
    node[right, font=\sffamily\normalsize, align=left] {$\ell$-redundancy};%$\operatorname{rank}(\bigcap_i Z(M_{f_i}))$\textbf{}
\draw[axis] (0,0) -- (0,5.05)
    node[above, font=\sffamily\normalsize, align=center] {typical fragment\\quantumness $\bar q$};

% Axis cues.
\node[font=\sffamily\normalsize, anchor=north] at (0.75,-0.10) {none};
\node[font=\sffamily\normalsize, anchor=north] at (6.85,-0.10) {large};
\node[font=\sffamily\normalsize, rotate=90, anchor=south] at (-0.20,0.88)
    {$\bar q=0$};
\node[font=\sffamily\normalsize, rotate=90, anchor=south] at (-0.20,4.15)
    {large $\bar q$};

\end{tikzpicture}